\documentclass[conference,a4paper]{IEEEtran}

\ifCLASSINFOpdf

\else

\fi
\usepackage{color,cite}
\usepackage{float}
\usepackage{soul}
\def\e{\begin{equation}}
\def\f{\end{equation}}

\def\_#1{{\bf #1}}
\def\=#1{\overline{\overline #1}}

\usepackage{graphicx}
\usepackage{placeins}
\usepackage{float}
\usepackage{subfigure}
\usepackage{amsmath} 

\renewcommand{\figurename}{Fig.}

\usepackage{color}
\usepackage{float}
\usepackage{color,soul}
\def\e{\begin{equation}}
\def\f{\end{equation}}

\usepackage{graphicx}
\usepackage{placeins}
\usepackage{float}
\usepackage{subfigure}
\usepackage[utf8]{inputenc} 
\usepackage[T1]{fontenc}
\usepackage{comment}

\newlength \figwidth
\makeatother

\IEEEoverridecommandlockouts
\IEEEoverridecommandlockouts\IEEEpubid{\makebox[\columnwidth]{ 978-1-6654-5975-
4/22~\copyright~2022 IEEE \hfill} \hspace{\columnsep}\makebox[\columnwidth]{ }}
\begin{document}

\title{Physically Consistent RIS: From Reradiation Mode
Optimization to Practical Realization}


\author{
\IEEEauthorblockN{Javad Shabanpour$^{\star}$, Constantin Simovski$^{\star}$, and Giovanni Geraci$^{\dagger\,\flat}$ \vspace{0.1cm}
}
\IEEEauthorblockA{$^{\star}$\emph{Aalto University, Finland} \quad\quad $^{\dagger}$\emph{Telefónica Research, Spain} \quad\quad $^{\flat}$\emph{Universitat Pompeu Fabra, Spain}
}

\thanks{This work was in part supported by H2020-MSCA-ITN-2020 META WIRELESS (Grant Agreement: 956256), by the Spanish Research Agency through grants PID2021-123999OB-I00, CEX2021-001195-M, and CNS2023-145384, by the UPF-Fractus Chair, and by the Spanish Ministry of Economic Affairs and Digital Transformation and the European Union NextGenerationEU through UNICO-5G I+D projects TSI-063000-2021-138 (SORUS-RIS) and TSI-063000-2021-59 (RISC-6G).}
}

\bstctlcite{IEEEexample:BSTcontrol}

\maketitle


\begin{abstract}

We propose a practical framework for designing a physically consistent reconfigurable intelligent surface (RIS) to overcome the inefficiency of the conventional phase gradient approach. For a section of Cape Town and across three different coverage enhancement scenarios, we optimize the amplitude of the RIS reradiation modes using Sionna ray tracing and a gradient-based learning technique. 
We then determine the required RIS surface/sheet impedance given the desired amplitudes for the reradiation modes, design the corresponding unitcells, and validate the performance through full-wave numerical simulations using CST Microwave Studio. 
We further validate our approach by fabricating a RIS using the parallel plate waveguide technique and conducting experimental measurements that align with our theoretical predictions.

\end{abstract}





\section{Introduction}

Reconfigurable Intelligent Surfaces (RIS) are an emerging technology being explored for future wireless networks. RIS are a type of metasurface that can boost the received signal power in weak and blind coverage zones \cite{basar2019wireless, shabanpour2022angular}. The European Telecommunications Standards Institute (ETSI) has established an industry specification group to coordinate RIS research and development, paving the way for the integration of RIS into future telecommunication standards \cite{sihlbom2022reconfigurable}.

A RIS is typically represented by a diagonal matrix, where each entry represents the phase shifts and reflection amplitudes corresponding to individual elements of the RIS \cite{wu2019intelligent}. The diagonal matrix model relies on a locally periodic approximation, ensuring that each unitcell scatters the incoming wave without affecting its neighbors. A comprehensive discussion on the reflection locality approximation, which is essential for applying the local periodicity assumption, can be found in \cite{shabanpour2024engineering}.

A new approach proposes moving beyond the diagonal matrix model to account for coupling effects between adjacent elements, resulting in a non-diagonal matrix where the off-diagonal elements represent interactions between different RIS elements \cite{li2022beyond}. Engineering such RIS presents significant challenges, as each unitcell must have its reflection phase independently controlled while considering the coupling effects. The coupling effects between adjacent unitcells mean that changing the reflection phase of one cell can affect its neighbors. This interdependence complicates the control algorithms and hardware design, as adjustments cannot be made in isolation; instead, they must consider the overall interaction network.

Both of the aforementioned approaches, which are based on rough approximations, fail to accurately model the electromagnetic (EM) behavior of RIS, especially for large deflection angles, which refer to the difference between incident and reflection angles. A physically consistent approach that integrates with ray-based models is proposed by \cite{degli2022reradiation, vitucci2024efficient}. These works present a realistic macroscopic model for evaluating multimode reradiation from a RIS, while considering power conservation. The model is designed for integration into forward ray tracing simulations, providing a computationally efficient solution for large-scale scenarios. This approach also supports both link- and system-level performance evaluations, even in realistic multipath propagation scenarios. However, the spatial modulation coefficient required for wave transformation by RIS, as presented in \cite{degli2022reradiation, vitucci2024efficient}, does not directly translate to the practical realization of unitcells.

To address this challenge, we propose a practical framework for designing a physically consistent RIS. Our approach accounts for the complex interactions between adjacent RIS elements, overcoming the limitations of the diagonal matrix model and accurately realizing the desired spatial modulation coefficient with high reradiation efficiency. Our contributions are summarized as follows:

\begin{itemize}

\item 
We optimize the amplitude of the RIS reradiation modes using Sionna ray tracing and a gradient-based learning technique. We perform this optimization for a section of Cape Town, considering three different coverage enhancement scenarios and a carrier frequency of 8 GHz.
\item 
We propose a framework for determining the required RIS surface/sheet impedance given the desired amplitudes for the reradiation modes. We design the corresponding unitcells and validate the performance through full-wave numerical simulations using CST Microwave Studio.
\item 
We further validate our proposed approach by fabricating a RIS prototype and conducting experimental measurements using the parallel plate waveguide technique that agree with our theoretical predictions.
\end{itemize}


    \renewcommand{\figurename}{Fig.}
  \begin{figure}
 	\centering
 	\includegraphics[width=\figwidth]{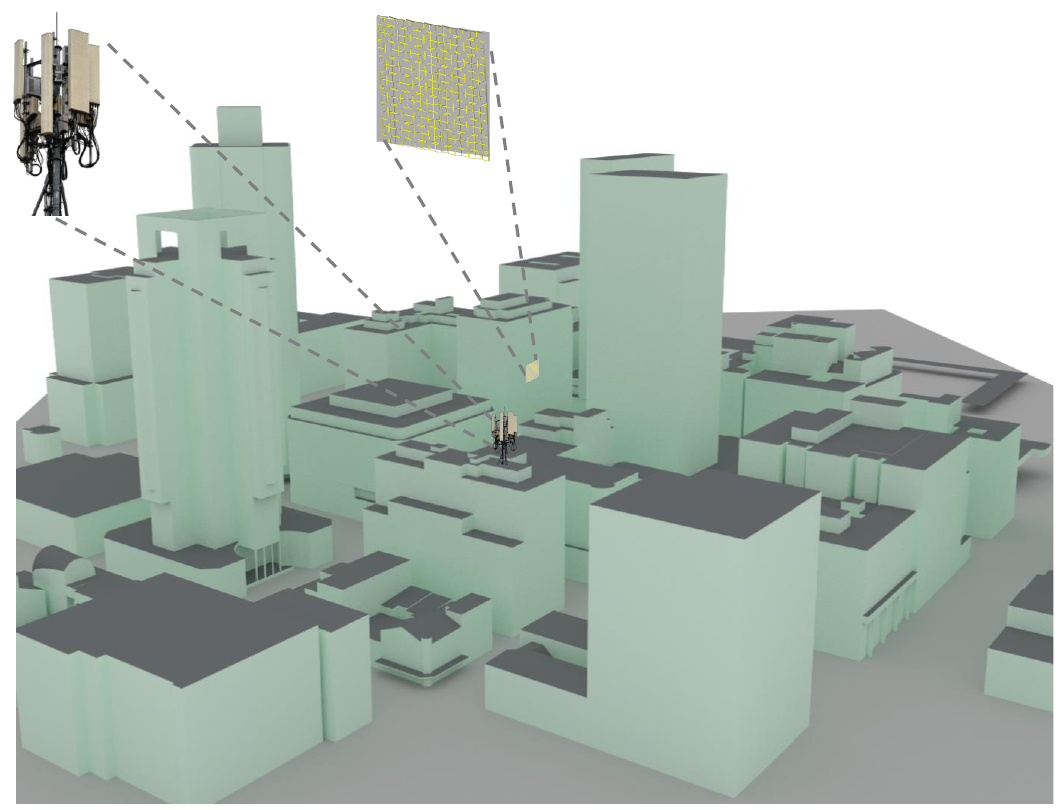}
 	\caption{Section of Cape Town, showing the location of base station and RIS.}
  \label{Fig1}
 \end{figure}



\section{Optimization of Reradiation Modes Amplitude}

In this section, we present our approach to optimize the amplitude of the RIS reradiation modes, i.e., the radiation eigenmodes of a periodically non-uniform metasurface.


\subsection{System Setup and Ray Tracing Simulations}

\subsubsection*{Frequency band}
Our target use case focuses on using RIS to improve service coverage in the FR3 band (7--24 GHz), which is envisioned to play a key role in 6G \cite{cui20236g}. 
Specifically, we consider a carrier frequency of 8\,GHz for all of our ray tracing simulations, numerical analysis, and measurements. 

\subsubsection*{Ray tracing tool}
We employ ray tracing because it offers a deterministic method for simulating the precise behavior of electromagnetic waves as they reflect, diffract, and scatter. In particular, we employ NVIDIA Sionna \cite{hoydis2023sionna}, a GPU-accelerated, open-source library based on TensorFlow. 
Sionna uses the RIS model from \cite{degli2022reradiation, vitucci2024efficient} in its ray tracer to calculate channel impulse responses and coverage maps. As a differentiable ray tracer for radio propagation modeling, Sionna allows for seamless integration with neural networks.

\subsubsection*{Scenario considered}
We select a section of Cape Town's city center, specifically the area defined by the coordinates $(-33.921, 18.4212)$ to $(-33.918, 18.4260)$, as illustrated in Fig.~\ref{Fig1}. This urban environment offers a complex scenario for evaluating the performance of the proposed physically consistent RIS model. A 3D representation of the selected area is constructed using OpenStreetMap data, including both terrain and building information. We configure the transmitter and receivers with single antennas featuring omnidirectional patterns and vertical polarization. Our setup can be adjusted for various numbers of elements, antenna patterns, and types of polarization.
The heights of the base station (BS) and the RIS are set to 33 and 40 meters, respectively. 
We assign material properties as follows: concrete for the streets, marble for the building walls, and metal for the rooftops. These material models are defined according to the ITU-R P.2040-2 recommendation \cite{series2015effects}. Our simulations consider a maximum of six ray bounces and trace ${10^7}$ random rays, accounting for line-of-sight propagation, reflection, scattering, and diffraction. The RIS dimensions are $82.5 \times 82.5$\,cm. Note that in physically consistent RIS modeling, conventional unitcells are not defined.

\subsubsection*{Low-coverage target areas}
To illustrate our proposed RIS design framework, we focus on two $10 \times 10$\,m target areas (\emph{Area1} and \emph{Area2}), highlighted in Fig.~\ref{Fig2}(a). The coverage map obtained without RIS deployment is depicted in Fig.~\ref{Fig2}(b), indicating that both target areas experience low coverage (around $-110$\,dB). Although we computed coverage maps for the entire area considered, for ease of visualization, Fig.~\ref{Fig2}(b)--(e) focus on a 
$150 \times 150$\,m area around the target areas.


\subsection{Reradiation Mode Optimization}

Given our focus on two target areas, we set the number of reradiation modes to two: \emph{mode1} and \emph{mode2}. Differentiable ray tracing allows us to optimize the amplitude of these modes using gradient-based learning techniques. Specifically, we utilize the Adam optimizer with a learning rate of $0.01$. Here, we provide three examples of coverage enhancement.

\subsubsection*{Equal signal strength}
In the first example, we address the fairness issue by ensuring that both target areas receive a similar level of signal strength. Initially, the amplitudes of the modes are set as trainable parameters, starting with values of $[0.01, 0.99]$. We then iteratively compute the coverage map in the target areas, allowing the optimizer to adjust the mode amplitudes to achieve the desired path gain. As shown in Fig.~\ref{Fig3}(a), after approximately $60$ iterations, the path gains for the two areas converge to $-93.6$\,dB, with the mode amplitudes stabilizing at $[0.41, 0.59]$.
Fig.~\ref{Fig2}(c) shows the RIS coverage gain compared to the scenario without RIS, demonstrating a significant improvement of about $25$\,dB around the target areas. The coverage improvement is more pronounced for area1 since it initially experienced poorer coverage.

\subsubsection*{Minimum guaranteed signal strength}
\begin{figure*}
	\centering
	\includegraphics[width=1.95\figwidth]{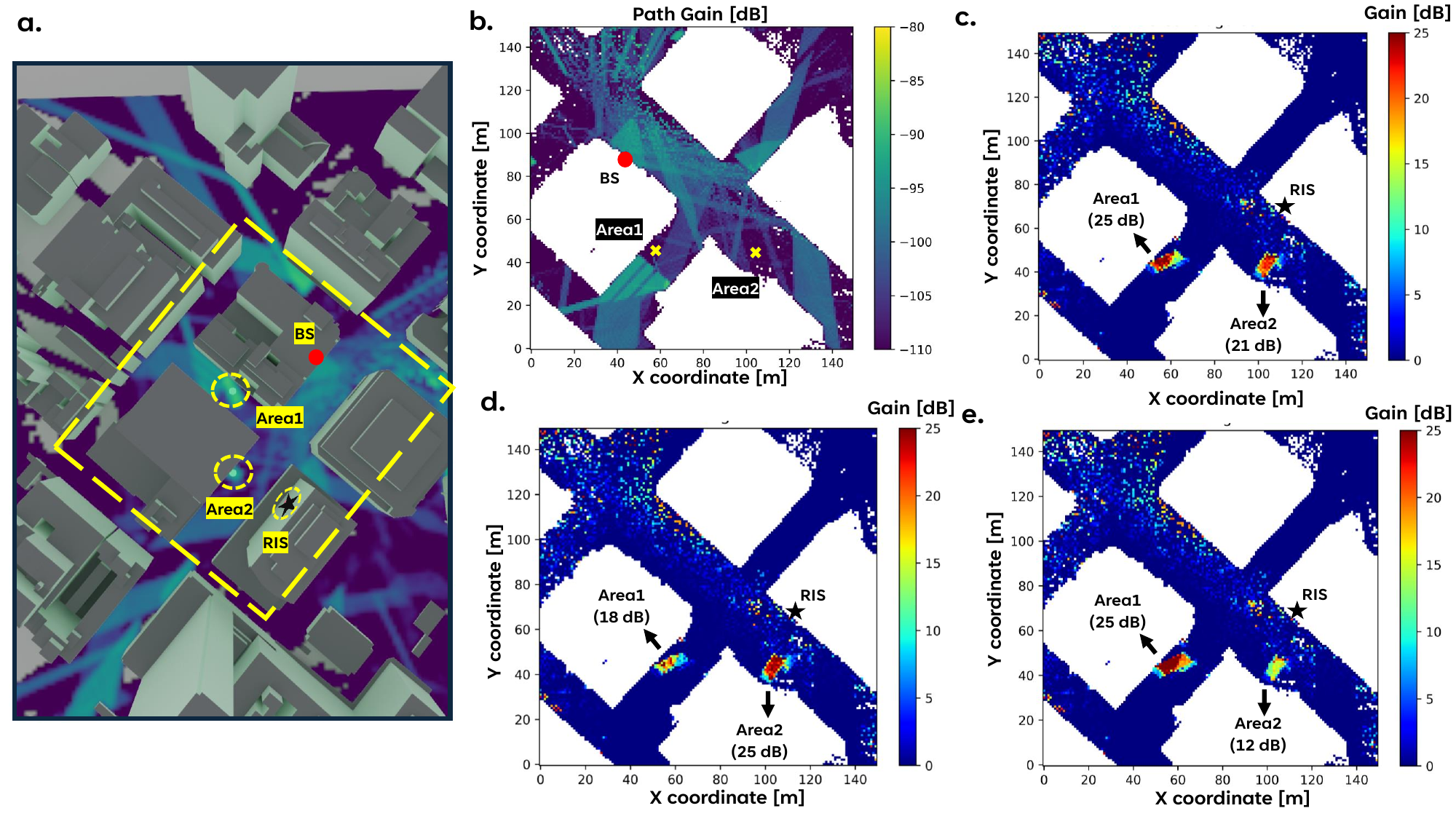}
	\caption{(a) Selected geographical area, showing the BS, RIS, and target areas.
(b) Coverage map without RIS.
(c) Coverage improvement with RIS optimized for equal signal strength. (d--e) Coverage improvement with RIS optimized for minimum guaranteed signal strength.}
\label{Fig2}
\end{figure*}

\begin{figure*}
	\centering
	\includegraphics[width=2\figwidth]{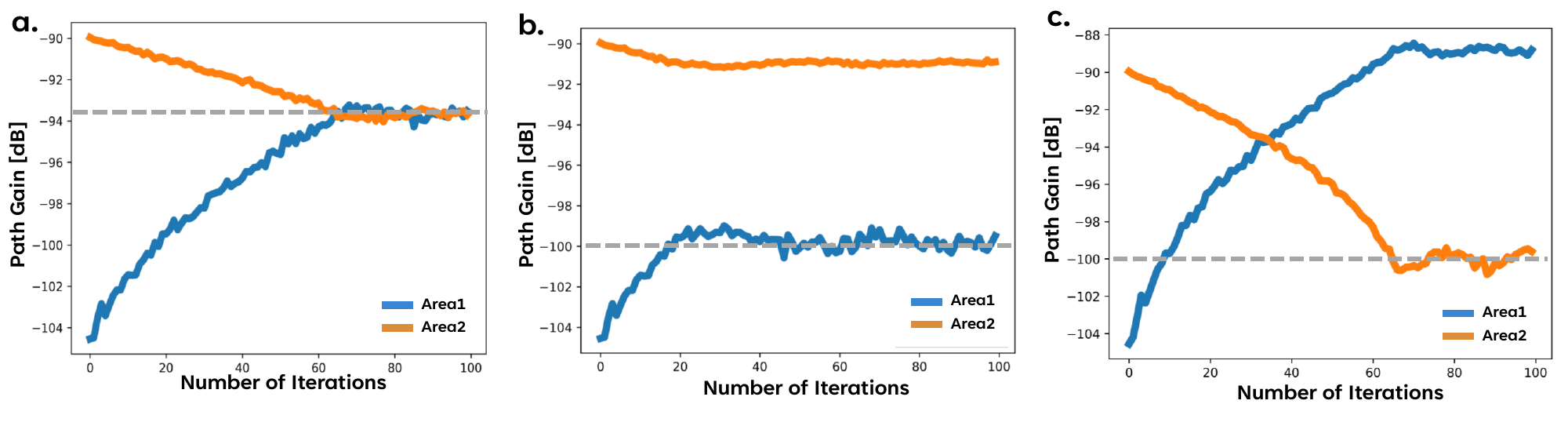}
	\caption{Gradient-based optimization of the RIS reradiation modes amplitude for: (a) equal signal strength on area1 and area2, (b) maximum signal strength on area2, while guaranteeing at least $-100$\,dB on area1; (c) maximum signal strength on area2 while guaranteeing at least $-100$\,dB on area1.}
\label{Fig3}
\end{figure*}

In the following two examples, we set the objective to ensure that the path gain for area1 (resp. area2) is at least $-100$\,dB while maximizing the average received signal power on area2 (resp. area1). Fig.~\ref{Fig3}(b) and Fig.~\ref{Fig3}(c) illustrate the corresponding evolution of the path gains based on the learned amplitude of reradiation modes during the training process. The results indicate that after approximately $50$ and $70$ iterations, the path gains on area1 and area2 converge to $-91.5$\,dB and $-89$\,dB, respectively, while maintaining a minimum path gain of $-100$\,dB on the other area. The mode amplitudes stabilize at $[0.19, 0.81]$ and $[0.71, 0.29]$, respectively. Fig.~\ref{Fig2}(d) and Fig.~\ref{Fig2}(e) show the RIS gain after deploying a physically consistent RIS with the aforementioned reradiation mode amplitudes. 


\renewcommand{\figurename}{Fig.}


\renewcommand{\figurename}{Fig.}


\section{Physically Consistent RIS Design Framework}

After determining the optimal amplitudes of the RIS reradiation modes, we now focus on their practical realization through surface impedance modeling.


\subsection{Preliminaries: From Impedance to Reradiation Modes}

The macroscopic spatial modulation coefficient $\Gamma (x',y')$, as defined in \cite{degli2022reradiation}, 
relates the amplitudes of the reflected Floquet harmonics to the incident field: ${{\bf{E}}_{\bf{r}}} = {{\bf{\Gamma }}_{{\rm{TE}}}} \cdot {{\bf{E}}_{\bf{i}}}$ for TE-polarized waves, and ${{\bf{H}}_{\bf{r}}} = {{\bf{\Gamma }}_{{\rm{TM}}}} \cdot {{\bf{H}}_{\bf{i}}}$ for TM-polarized waves. For the practical realization of $\Gamma (x',y')$, one can employ a method based on mode matching, as detailed in \cite{kosulnikov2023discrete}. The desired RIS behavior is then characterized through a surface impedance matrix. 



Surface impedance is related to the spatial modulation coefficient for TE and TM polarization as follows: 
\begin{equation}
{{\bf{\Gamma }}_{\textrm{TE}}} = \frac{{{{\bf{Y}}_0} - {{\bf{Y}}_{\textrm{tot}}}}}{{{{\bf{Y}}_0} + {{\bf{Y}}_{\textrm{tot}}}}},\,\quad{{\bf{\Gamma }}_{\textrm{TM}}} = \frac{{{{\bf{Z}}_0} - {{\bf{Z}}_{\textrm{tot}}}}}{{{{\bf{Z}}_0} + {{\bf{Z}}_{\textrm{tot}}}}}
\label{Eq1}
\end{equation}
where ${{\mathbf{Z}_0}}$ is the matrix of free space impedance referred to the metasurface plane, consisting of identical diagonal elements, and ${{\mathbf{Y}_0}}$ is a similar matrix of free-space admittance. By accurately calculating the total impedance $\mathbf{{Z_{\textrm{tot}}}}$ of the metasurface, the spatial modulation coefficient of the given reradiation mode can be determined. 

In the surface impedance model, the metasurface is assumed to consist of an impedance sheet on top of a grounded substrate. Let ${Z_s}(x)$ be the sheet impedance of a metasurface with period $D$. This can be expressed as a Fourier series:
\begin{equation}
{Z_s}(x) = \sum\limits_{m =  - \infty }^{ + \infty } {{a_m}{e^{ - jm{\beta _M}x}}},
\label{Eq2}
\end{equation} 
where ${\beta _M} = 2\pi /D$ is the spatial modulation frequency of the surface impedance, and ${a_m}$ are the Fourier coefficients. 
The reflection angle of the $n$-th propagating mode can be expressed as \cite{shabanpour2024engineering}:
\begin{equation}
\sin {\theta _{r,n}} = \sin {\theta _i} - \frac{{n\lambda }}{D},
\label{Eq3}
\end{equation}
where $\lambda$ is the wavelength.
Since the metasurface is periodically modulated, it can generate an infinite number of reradiation modes. Considering the equivalent circuit model of the metasurface at the supercell scale, the corresponding voltage $V_s$ and current $I_s$ in the transmission line can be expressed as: 
\begin{equation}
{I_s}(x) = \sum\limits_{n =  - \infty }^{ + \infty } {i_s^n{e^{ - j{k_{xn}}x}}} ,\,\,\,\,\,{V_s}(x) = \sum\limits_{n =  - \infty }^{ + \infty } {v_s^n{e^{ - j{k_{xn}}x}}}.
\label{Eq4}
\end{equation}
The grounded substrate is
modeled as a shorted transmission line, with its length equal to the substrate thickness. By applying Ohm's law and substituting (\ref{Eq2}) into (\ref{Eq4}): 
\begin{equation}
\sum\limits_{n =  - \infty }^{ + \infty } {v_s^n{e^{ - j{k_{xn}}x}}}  = \sum\limits_{m =  - \infty }^{ + \infty } {\sum\limits_{n =  - \infty }^{ + \infty } {{a_m}i_s^n{e^{ - j{k_{x,n + m}}x}}} }.
\label{Eq5}
\end{equation}
After transforming from $n$ to $n-m$ on the right-hand side of (\ref{Eq5}) and considering a finite number of Floquet harmonics from $-N$ to $N$, we obtain: 
\begin{equation}
\left( {\begin{array}{*{20}{c}}
{v_s^{ - N}}\\
{v_s^{1 - N}}\\
 \vdots \\
{v_s^{ + N}}
\end{array}} \right) = \left( {\begin{array}{*{20}{c}}
{{a_0}}&{{a_{ - 1}}}& \cdots &{{a_{ - 2N}}}\\
{{a_1}}&{{a_0}}& \cdots &{{a_{1 - 2N}}}\\
 \vdots & \vdots & \ddots & \vdots \\
{{a_{2N}}}&{{a_{2N - 1}}}& \cdots &{{a_0}}
\end{array}} \right)\left( {\begin{array}{*{20}{c}}
{i_s^{ - N}}\\
{i_s^{1 - N}}\\
 \vdots \\
{i_s^{ + N}}
\end{array}} \right)\,
\label{Eq6}
\end{equation}
where each voltage harmonic is interconnected with all current harmonics due to mode coupling. 

Similar to sheet impedance, the total surface impedance of the metasurface is defined as a parallel connection of the discretized sheet impedance matrix 
$\mathbf{Z}_{\textrm{s}}$ and the matrix of the uniform impedance of the grounded dielectric substrate \cite{kosulnikov2023discrete}. 
Once the total surface impedance is determined, all scattered modes for a given incidence can be calculated using the reflection matrix in (\ref{Eq1}).

\renewcommand{\figurename}{Fig.}
\begin{figure}[ht]
	\centering
	\includegraphics[width=\figwidth]{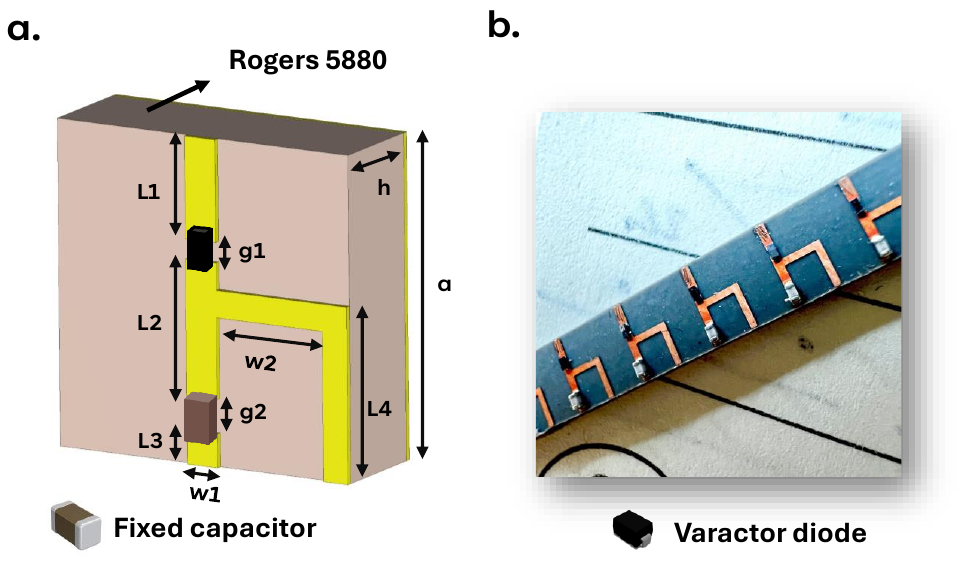}
	\caption{(a) Geometry of the unitcell designed via CST, with values provided in Table~I. (b) Fabricated samples of unitcells as detailed in Section~V.}
\label{Fig4}
\end{figure}
\subsection{Unitcell Design Framework}

In Section~III-A, we reviewed the calculation of scattered harmonics for a set of discrete impedance sheets and illuminating waves. However, designing a metasurface is an \emph{inverse problem}, where one seeks to determine the appropriate sheet impedance by specifying the incident and reflected electric fields. Our proposed approach to this inverse problem consists of three steps, detailed as follows.

\subsubsection{Determining the supercell period}
Based on diffraction grating theory as shown in (\ref{Eq3}), we determine the period of one supercell based on the incident and desired reradiation angles.
The supercell is then discretized into $K$ unitcells where ${Z_{1 \ldots K}}$ represents the discrete sheet impedance for the unitcells.

\subsubsection{Specifying incident and reflected fields}
Considering plane wave illumination, the incident electric field can be expressed as: ${\mathbf{E}_i} = {[0,\ldots, 0,1,0,\ldots,0]^T}$. The reflected electric field vector can be expressed as ${{\bf{E}}_{\bf{r}}} = {[0, \ldots {A_{{\rm{mode}}1}},0,{A_{{\rm{mode2}}}}, \ldots 0]^T}$, where ${A_{{\rm{mode}}1}}$ and ${A_{{\rm{mode}}2}}$ represent the amplitudes of the reradiation modes (harmonics $n=\pm1$). In Section~II, we observed the effect of optimizing these mode amplitudes on the coverage map via ray tracing. Note that the specular reflection amplitude (harmonic $n=0$) is set to zero.

\subsubsection{Optimizing the sheet impedance}
We use Matlab's mathematical optimization capabilities to identify the optimal values for the impedance sheets. In each trial, the optimizer assumes an explicit array ${Z_{1 \ldots K}}$ and calculates the amplitude of the reflected harmonics ${A_{\rm cal}}$ using (\ref{Eq4})--(\ref{Eq6}). By defining the cost function as $F({Z_{1 \ldots K}}) = \left| {{A_{{\rm{cal}}}} - {A_{{\rm{modes}}}}} \right|$ and employing the MultiStart and Fmincon optimization algorithms, we search for the matrix ${Z_{1 \ldots K}}$ that minimizes this cost function. The optimal ${Z_{1...K}}$ is then realized by designing the unitcells as follows.%
\footnote{Our framework can be extended to dual polarization, yielding a different surface impedance for each desired spatial modulation coefficient as per (\ref{Eq1}).}
%


\subsubsection*{Designing the unitcell}

The goal of designing the unitcells is to ensure they cover a specific range of sheet reactance, which is the imaginary part of the sheet impedance. Our analytical calculations in Matlab indicate that a reactance range from 
$-70$ to $-185$ Ohms is sufficient 
to realize the desired impedance values, as later shown in (7)--(9).
Fig.~\ref{Fig4}(a) illustrates the detailed geometry of the designed unitcell, with all parameters specified in Table~\ref{tab:geometry}. 
Fig.~\ref{Fig4}(b) shows samples of fabricated unitcells, consisting of a microstrip wire on a metal-backed RT5880 substrate, loaded with a varactor diode and an additional capacitor (see Section~V).

\begin{table}[h]
\centering
\caption{Geometry of the designed unitcell ($w1=g2=L3$).}
\begin{tabular}{l c c c c c c c c}
\hline
 \textbf{Parameter} & a & h & L1 & L2 & L4 & w2 & g1 & L3 \\
\hline
 \textbf{Value [mm]} & 4.8 & 1.575 & 1.5 & 2 & 2.6 & 2.15 & 0.3 & 0.5 \\
\hline
\end{tabular}
\label{tab:geometry}
\end{table}

\noindent The first gap ${g_1}$ of the microstrip wire, shown in Fig.~\ref{Fig4}(a), is filled with a hyper-abrupt varactor diode MAVR-000120-14110P. This varactor can operate at frequencies up to $70$\,GHz, offering low series resistance and a capacitance range of $0.1$--$1.1$\,pF. The constant capacitor, which loads the second gap ${g_2}$, is GJM1555C1HR15RB12D. This capacitor is chosen for its fine tolerance of $\pm 0.03$ pF and has a nominal capacitance of 0.15 pF.


\renewcommand{\figurename}{Fig.}
\begin{figure*}
	\centering
	\includegraphics[width=2\figwidth]{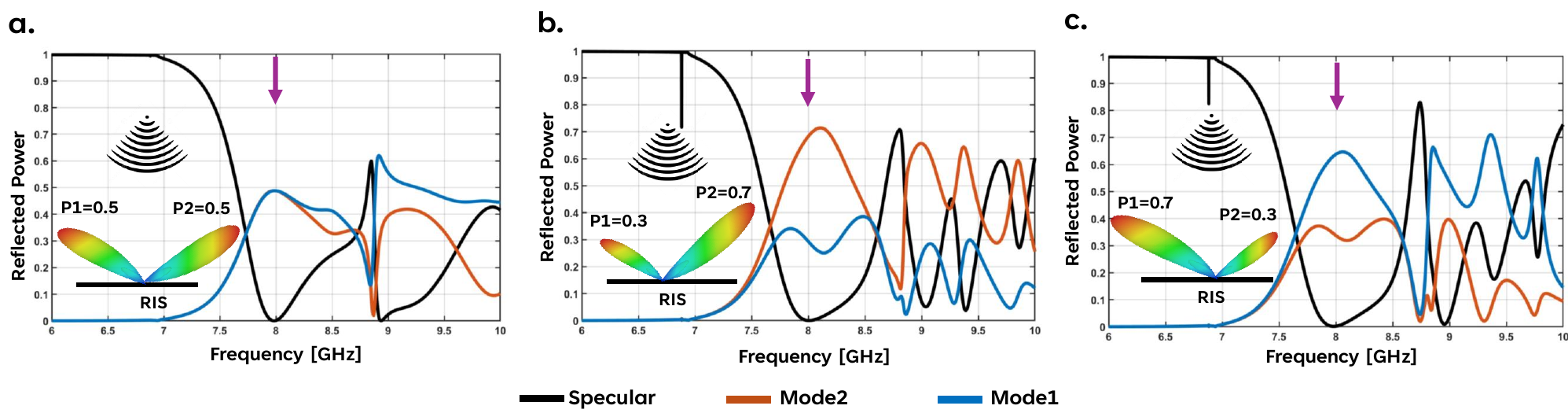}
	\caption{Full-wave numerical simulation of the reradiation mode power obtained for the discrete sheet reactance values ${Z_{1 \textrm{--} 9}^{\textrm{a}}}$, ${Z_{1 \textrm{--} 9}^{\textrm{b}}}$, and ${Z_{1 \textrm{--} 9}^{\textrm{c}}}$.} 
\label{Fig5}
\end{figure*}


\subsection{Numerical Validation}

With the unitcell design complete and the optimizer configured, we are prepared to realize any spatial modulation coefficient $\Gamma (x',y')$. 
We utilize CST Microwave Studio to calculate the reradiations from the RIS by applying periodic boundary conditions in the x- and y-directions. The operating frequency is 8 GHz, and the width of each unitcell is $P = 4.8$\,mm. The illumination is a plane wave with normal incidence. The number of unitcells in the supercell is fixed at $K=9$. Based on (\ref{Eq3}), there would be three reradiation modes toward ${\theta _r} = {60^{\circ}}$ (mode1), ${\theta _r} = {-60^{\circ}}$ (mode2), and specular reflection (mode0). To validate the process, we consider three scenarios, each respectively with the following mode amplitudes: $[0.5, 0.5]$, $[0.3, 0.7]$, and $[0.7, 0.3]$.%
\footnote{For simplicity, we demonstrate our approach using mode amplitudes that are slightly different from the values obtained via ray tracing, i.e., $[0.5, 0.5]$, $[0.3, 0.7]$, and $[0.7, 0.3]$ versus $[0.41, 0.59]$, $[0.19, 0.81]$, and $[0.71, 0.29]$. However, any amplitude values can be realized with the proposed approach.}

In the first scenario, our goal is to eliminate the specular reflection and distribute the power equally between the two modes. The optimized values of the discrete sheet impedance matrix for this example are as follows:
\begin{equation}
\begin{aligned}
{Z_{1 \textrm{--} 9}^{\textrm{a}}} = [&-50j, -98j, -190j, -110j, -145j,\\
&-185j, -66j, -87j, -139j]
\end{aligned}
\end{equation}
These impedance values are then configured in our unit cells in CST Microwave Studio, where we carry out full-wave numerical simulations. 
Fig.~\ref{Fig5}(a) shows the reradiation mode power obtained for the discrete sheet reactance values ${Z_{1 \textrm{--} 9}^{\textrm{a}}}$. These align closely with our theoretical predictions and confirm that the incident plane wave is nearly fully reflected into the desired reradiation modes with the predefined reflection amplitudes. The purple arrow in the figure indicates the operating frequency of 8 GHz.

In the second and third examples, we direct 30\% of the reradiated power to mode1 and 70\% to mode2, and vice versa, respectively. The respective optimized discrete sheet reactance values are as follows:
\begin{equation}
\begin{aligned}
{Z_{1 \textrm{--} 9}^{\textrm{b}}} = [&-85j, -90j, -185j, -146j, -55j,\\
&-167j, -185j, -77j, -101j]
\end{aligned}
\end{equation}
\begin{equation}
\begin{aligned}
{Z_{1 \textrm{--} 9}^{\textrm{c}}} = [&-140j, -70j, -189j, -85j, -1555j,\\
&-155j, -190j, -120j, -75j].
\end{aligned}
\end{equation}

\noindent Fig.~\ref{Fig5}(b) and Fig.~\ref{Fig5}(c) illustrate the reradiation modes power obtained for the discrete sheet reactance values ${Z_{1 \textrm{--} 9}^{\textrm{b}}}$ and ${Z_{1 \textrm{--} 9}^{\textrm{c}}}$, respectively. The specular reflection is nearly zero at 8 GHz and the numerical results show excellent agreement with our predefined goals.  


\section{Experimental Validation}

\begin{figure*}
	\centering
	\includegraphics[width=1.8\figwidth]{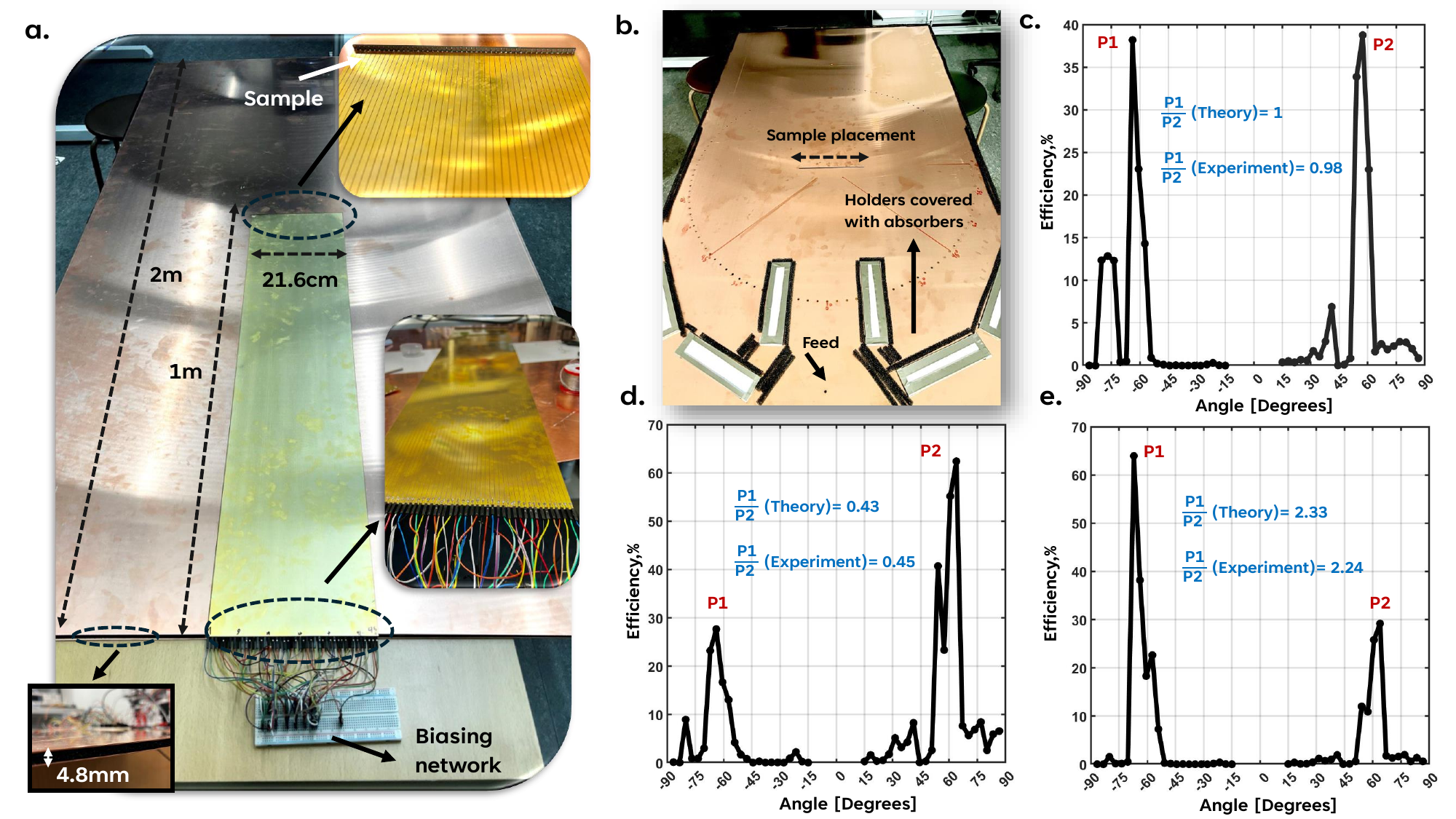}
	\caption{(a) Experimental setup with insets showing: the gap between the plates (bottom left), the sample soldered to the strips of the printed circuit board (top right), and jumper wires connected to the other side of the printed circuit board (center right).
(b) Lower plate of the parallel plate waveguide with holes drilled for inserting the receiving probe.
(c--e) Measured mode reradiation efficiency corresponding to the three examples presented in Fig.~\ref{Fig5}(a--c), respectively.}
\label{Fig6}
\end{figure*}

To further validate the proposed design framework for a physically consistent RIS, we conducted three experiments corresponding to the examples presented in Fig.~\ref{Fig5}(a-c), each with different reradiation mode power allocations.
In our experiments, the incident wave is a plane wave with vertical polarization, consistent with the ray-tracing setup.
Since the RIS is uniform in the y-direction, we performed the measurements inside a parallel plate waveguide (PPW) \cite{asadchy2015functional}.
The PPW effectively captures both near-field and far-field behavior of 2D metasurfaces that are homogeneous in one direction. This approach significantly simplifies the metasurface design process, as it requires fabricating only a single row of the surface.
Based on image theory, our 1D array mimics an infinite metasurface in the y-direction and comprises 45 unitcells in the x-direction, as seen in Fig.~\ref{Fig6}(a). By adjusting the bias voltage of each varactor, we achieve the impedance matrices 
${Z_{1 \ldots K}}$ for each scenario.


\renewcommand{\figurename}{Fig.}


Fig.~\ref{Fig6}(a) shows the complete setup, which consists of two copper plates, each sized $1 \times 2$ m. Fig.~\ref{Fig6}(b) illustrates the lower copper plate with a circular array of 3 mm diameter perforations arranged on a radius of 0.49 meters to accommodate the measurement probe. The transmitting probe, fixed at location Feed 1, excites a quasi-TEM wave in the waveguide. The circular array of holes allows us to measure the complex transmission coefficient $S_{21}$ as a function of the scattering angle $\theta$ in $3^{\circ}$ steps using a vector network analyzer. For biasing implementation, we created a cutaway in the top copper plate and inserted a PCB with parallel strips on its bottom surface, separated by small gaps (as shown in the inset of Fig.~\ref{Fig6}(a)). This innovative biasing configuration allows for the separation of DC and AC operations. At zero frequency, each varactor is independently adjusted by its DC voltage, while at AC frequencies, the microstrip wires of each unitcell are effectively connected to both metal plates of the waveguide. 

We conducted measurements for three reradiation mode power allocations based on the numerical simulations in Fig.~\ref{Fig5}(a-c). Using the sheet reactance matrices $Z_{1-9}$ obtained in Section III-C, we calculated the necessary varactor capacitances $C_v$. The corresponding reverse DC voltages were then determined from the varactor datasheet and applied using the specified biasing system. Fig.~\ref{Fig6}(c-e) display the efficiency after the required normalizations. The results show that the theoretical predictions for the mode power allocations align well with the measurements. P1 and P2 represent the peak power for the first and second reradiation modes at reflection angles 
${\theta _r} = {60^{\circ}}$ and ${\theta _r} = {-60^{\circ}}$, respectively. We note that the remaining power is either dissipated due to varactor losses or falls into blind spots. 
From the curves, it can be seen that the capacitor and varactor losses are approximately 10\%, which is considered within an acceptable range.


\section{Conclusion}

We proposed a new framework for the practical realization of physically consistent RIS. We optimized the RIS reradiation modes using Sionna ray tracing and a gradient-based learning technique for a section of Cape Town and three different coverage enhancement scenarios. To translate the resulting spatial modulation coefficient into a practical unitcell realization, we proposed a mode-matching approach that enables realistic surface impedance design using varactor diodes. We validated the designed RIS through full-wave simulations conducted in CST Microwave Studio. Additionally, we fabricated a RIS using the parallel plate waveguide technique to further validate our approach. Our proposed framework overcomes the limitations of current approaches based on diagonal matrices or phased arrays, achieving the desired RIS behavior with high accuracy and efficiency.



\bibliography{main}   
\bibliographystyle{IEEEtran}
\end{document}